# The emergence of giant multicaloric phenomena near room temperature

Tapas Samanta[1*], Pol Lloveras[2], Ahmad Us Saleheen[1], Daniel L. Lepkowski[1], Emily Kramer[1], Igor Dubenko[3], Philip W. Adams[1], David P. Young[1], Maria Barrio[2], Josep Ll. Tamarit[2], Naushad Ali[3] and Shane Stadler[1]

[1]Department of Physics & Astronomy, Louisiana State University, Baton Rouge, LA 70803 USA
[2]Departament de Física, ETSEIB, Universitat Politècnica de Catalunya, Diagonal 647, Barcelona, 08028 Catalonia, Spain
[3]Department of Physics, Southern Illinois University, Carbondale, IL 62901 USA

Caloric responses – temperature changes – can be induced in solid-state materials by applying external stimuli such as stress, pressure, and electric and magnetic fields. The magnetic-field-stimulated response is called the magnetocaloric effect, and materials that exhibit this property have long been sought for applications in room temperature magnetic cooling due to their potentially superior efficiency and low impact on the environment. Other solid-state caloric phenomena are less developed, but are likewise under intense investigation. Here we introduce a new material that not only displays giant barocaloric (hydrostatic-pressure-induced) properties, but also a large magnetocaloric response near room temperature. It is unprecedented that two caloric effects of such extreme magnitude occur in the same material and at the same temperature. These effects originate from a magnetostructural transition and a magneto-volume (magnetostriction) effect where the volume change is large enough to force the system from a localized ordered state into an itinerant paramagnetic state.

[*]Correspondence to: tapas.sinp@gmail.com

The solid-state caloric effect is defined as a change of entropy or temperature in a material in response to an isothermal or adiabatic variation of an externally applied parameter such as magnetic field, stress, pressure, or electric field. Materials that exhibit solid-state caloric phenomena could form the basis of more environmentally-friendly and efficient alternatives to conventional, vapor-compression-based cooling devices. Significant progress has been made over the past decade in the development of caloric effects induced by magnetic field, i.e., the magnetocaloric effect (MCE), which has been the most promising applied caloric phenomenon until now.[1-5] However, the less-studied pressure-induced effect, called the barocaloric effect (BCE), has shown promise after recent discoveries of the phenomenon in solid-state materials.[6-10] The barocaloric effect is expected to occur in any solid material that undergoes a pressure-induced volume change. In turn, as predicted theoretically, the most likely candidates to show both large baro- and magneto-caloric effects (i.e., multicaloric effects) are materials that undergo a first-order magnetic transition (FOMT) with a simultaneous large change in volume.[11] The combined caloric effects in a single material will provide a way to improve the effectiveness and efficiency of environmentally friendly solid-state cooling. Here we report the discovery of giant baro- and magneto-caloric effects near room temperature in the MnNiSi-based material, $(MnNiSi)_{1-x}(FeCoGe)_x$ ($x = 0.38$). The isothermal entropy changes in this compound were 73.7 and −58.2 J kg$^{-1}$ K$^{-1}$ for change in hydrostatic pressure of $P = 2.69$ kbar and change in magnetic field of $\mu_0 H = 5$ T, respectively.

There are two classes of magnetic materials that exhibit large volume changes when undergoing a FOMT. In the first, which includes localized magnetic systems such as $Gd_5Si_2Ge_2$ (ref. 12) and Ni-Mn-In alloys[6], magnetocrystalline coupling is responsible for the volume change as the system undergoes a magnetostructural transition (MST). Instead, in itinerant magnetic systems such as La(Fe,Co,Si)$_{13}$ (ref. 13), the volume change is connected to a spontaneous magnetostriction correlated with a change in amplitude of the local moment at the transition. Here we show that a MnNiSi-based system possesses both of these effects: a magnetostructural transition (localized magnetism) followed by a pressure-induced magnetostriction (itinerant magnetism). The result is an unprecedentedly large barocaloric effect in a system that also exhibits giant magnetocaloric effects.

In recent years, Mn-based ternary compounds with the formula Mn$TX$ ($T$ = Ni, Co and $X$ = Si, Ge) have attracted considerable attention because of their pronounced magnetocaloric properties near room temperature. In these compounds, the effects originate from a magnetostructural transition with a large volume change of about 3-4%.[14-19] In their stoichiometric forms, the materials in this family undergo a second-order magnetic transition with variety of magnetic structures in the ordered state (depending on composition), such as collinear ferromagnetic (e.g., in MnCoGe and MnNiSi), spiral antiferromagnetic (in MnNiGe), and noncollinear helical antiferromagnetic (in MnCoSi) structures. In the paramagnetic state, all of the compounds undergo a martensitic structural transition from a low-temperature orthorhombic TiNiSi-type structure to a high-temperature hexagonal $Ni_2In$-type structure accompanied by a large change in volume. In order to obtain large caloric effects, it is crucial that the magnetic and structural transitions be coupled. This can often be accomplished by varying the chemical composition[14], changing the stoichiometry[15,16], isostructurally substituting[17-19], or by applying pressure[20]. Using these tuning strategies, coupled magnetostructural transitions have been realized in many of these materials near room temperature, along with giant magnetocaloric effects bolstered by a large additional structural contribution to the total isothermal entropy change. Generally, the structural transition responds more quickly to external stimuli than the magnetic transition. Consequently, when magnetocrystalline coupling is present, the magnetostructural transition proceeds at the rate of the purely structural transition. It is important for applications that the MST remains coupled over a wide temperature range that spans room temperature.

In our recent study of (MnNiSi)$_{1-x}$(FeCoGe)$_x$, we found that isostructural substitution results in a coupled MST from a ferromagnetic to a paramagnetic state, and that the transition temperature shifts drastically with applied hydrostatic pressure (up to 10 K/kbar).[19] Moreover, the transition remains coupled with applied pressures greater than 5 kbar. This results in a large MCE over a large, pressure-widened temperature range (from 235-355 K). In this case, the MCE is due to a large structural entropy change from a high-volume, low-symmetry orthorhombic to a low-volume, high-symmetry hexagonal phase. This large volume change and the shift of the MST to lower temperature with pressure inspired us to conduct a study of the barocaloric properties of the (MnNiSi)$_{1-x}$(FeCoGe)$_x$ system. Here we report a giant, inverse barocaloric effect that spans a wide temperature range and corresponds to a large refrigeration capacity (1508 J kg$^{-1}$ ≡ 11 J cm$^{-3}$ for $P$ = 2.69 kbar). This extreme barocaloric effect, together with its giant magnetocaloric effect, makes (MnNiSi)$_{1-x}$(FeCoGe)$_x$ a giant multicaloric material.

**Results**

The large caloric effects occur over a wide temperature range (255 K < $T$ < 338 K) by varying the concentration (0.38 < $x$ < 0.41). However, since the magnetostructural transition shifts to lower temperature with pressure for this study, we selected a composition ($x$ = 0.38) with the phase transition above room temperature ($T_M$ = 338 K at $P$ = 0). The temperature dependent magnetization data for 0.1 and 5 T applied magnetic fields and for different applied hydrostatic pressures (under $\mu_0 H$ = 0.1 T) are shown in Fig. 1a. Calorimetric curves for different hydrostatic pressures are shown in Fig. 1b. The opposite shifts in the magnetostructural transition temperature by applying pressure (negative shift) and magnetic field (positive shift) are associated with the stabilization of the high-temperature, low-volume hexagonal phase with pressure and low-temperature ferromagnetic orthorhombic phase with magnetic field, respectively. This pressure- and magnetic-field-induced cross-correlation of magnetocrystalline coupling is responsible for an inverse BCE and a conventional MCE.

Due to the large difference in unit cell volume and, hence, interatomic spacing, a larger magnetic moment has been reported for the orthorhombic structure relative to that of the hexagonal phase.[21] This is connected to the localized, narrower $3d$ band in the orthorhombic structure compared to that of the hexagonal structure, resulting in itinerant magnetic character in the latter due to the shorter bond length.[22,23] Note that the saturation magnetization is independent of pressure in Fig. 1c. This suggests that localized magnetism persists in the low-temperature orthorhombic phase with application of pressure. In contrast, the deviation of the inverse susceptibility ($1/\chi$) from a linear Curie-Weiss behavior in Fig. 1d indicates that the system has an itinerant magnetic character in the hexagonal paramagnetic phase.

The endothermic peaks in the heating cycle of the differential scanning calorimetry (DSC) measurements (Fig. 1b) are a measure of the latent heat of the transition. The latent heat at ambient pressure, calculated as area beneath the heating curve, is equal to 20.8 kJ kg$^{-1}$ (151.8 J cm$^{-3}$) and corresponds to a transition entropy change of $\Delta S_t$ = 61.8 J kg$^{-1}$ K$^{-1}$ (451.1 mJ cm$^{-3}$ K$^{-1}$). The pressure dependent transition entropy changes as a function of temperature, as calculated from the calorimetric curves, are shown in Fig. 2. The transition temperature shifts to lower temperature with increasing hydrostatic pressure at a rate of d$T_M$/d$P$ = −7.5 K kbar$^{-1}$ (see the inset of Fig. 2). Interestingly, the transition entropy increases from about 61.8 to 73.7 J kg$^{-1}$ K$^{-1}$ (11.9 J kg$^{-1}$ K$^{-1}$ ≡ 86.9.1 mJ cm$^{-3}$ K$^{-1}$ as $P$ goes from 0 to 2.69 kbar) with increasing pressure.

The temperature dependence of the pressure-induced entropy change ($\Delta S_{BCE}$) was calculated from the difference between the transition entropy curves (Fig. 2) at pressure $P$ ($\Delta S_t(T,P)$) and $P$ = 0 ($\Delta S_t(T,0)$). The results shown in Fig. 3a indicate a giant *inverse* barocaloric effect near room temperature. Notably, $\Delta S_{BCE}$ increases with increasing pressure up to the highest applied pressure, and its value is significantly larger than the transition entropy at ambient pressure ($\Delta S_t(T,0) = 61.8$ J kg$^{-1}$ K$^{-1}$). The maximum value of $\Delta S_{BCE}$ at $P = 2.69$ kbar is 73.7 J kg$^{-1}$ K$^{-1}$ with a reversible change of 56.9 J kg$^{-1}$ K$^{-1}$. For devices it is desirable to have the largest possible entropy change in the smallest possible volume and, therefore, it is helpful compare materials based on entropy density.[4] The barocaloric effect in terms of entropy density for (MnNiSi)$_{1-x}$(FeCoGe)$_x$ with $x = 0.38$ (538 mJ cm$^{-3}$ K$^{-1}$ for $P = 2.69$ kbar) outperforms all other reported materials (see Table 1). The relative cooling power (RCP = $|\Delta S_{BCE}^{max} \times$(FWHM of $\Delta S_{BCE}(T))|$) is another important parameter to estimate the applicability of a material for solid-state cooling. Calculated from the $\Delta S_{BCE}(T)$ curves in Fig. 3a, the RCP is 1508 J kg$^{-1}$ (11 J cm$^{-3}$) for $P = 2.69$ kbar (see the inset of Fig. 3a).

The important element that sets (MnNiSi)$_{1-x}$(FeCoGe)$_x$ as a multicaloric system is that it also exhibits a giant magnetocaloric (magnetic-field-induced) effect at the same transition responsible for the giant BCE. The field-induced isothermal entropy change ($\Delta S_{MCE}$) is $-58.2$ J kg$^{-1}$ K$^{-1}$ ($-424.9$ mJ cm$^{-3}$ K$^{-1}$) for a field change of $\mu_0 H = 5$ T (Fig. 3b) as calculated from magnetization isotherms using a Maxwell relation (see Methods). The value of $\Delta S_{MCE}$ was also determined by employing the Clausius-Clapeyron equation following ref. 26, [$\frac{\Delta S}{\Delta M} = \frac{\mu_0 dH}{dT} \rightarrow \Delta S \sim (\Delta M/\Delta T)\mu_0\Delta H$] resulting in $\Delta S_{MCE} = -56.6$ J kg$^{-1}$ K$^{-1}$ for $\mu_0 H = 5$ T ($\Delta M = -67.9$ A m$^2$ kg$^{-1}$ and $\Delta T = 6$ K for $\mu_0 H = 5$ T). This value is comparable to or larger than those reported for the best giant magnetocaloric materials known so far (see Table 1).

In order to fully characterize the barocaloric effects, the adiabatic temperature change must also be known. We have measured the pressure-induced adiabatic temperature change ($\Delta T_{BCE}$) upon compression and decompression, and also estimated it using heat capacity data. As an example, $\Delta T_{BCE}$ was estimated in ferroelectric (NH$_4$)$_2$SO$_4$ (ref. 8) using heat capacity data, giving a value of $\Delta T_{BCE} = 8.0$ K for $P = 1.0$ kbar. Using the same method for (MnNiSi)$_{1-}$ $_x$(FeCoGe)$_x$ gives a temperature change of $\Delta T_{BCE} = \frac{T}{c}|\Delta S|_{BCE} \sim 33$ K (see Table 1), using the reversible $\Delta S_{BCE} = 56.9$ J kg$^{-1}$ K$^{-1}$ at 320 K for $P = 2.69$ kbar and the specific heat capacity $c = 550$ J kg$^{-1}$ K$^{-1}$ (see the inset of Fig. 4a).

Another estimation of the adiabatic temperature change with pressure is the pressure-induced shift of the phase transition temperature. Returning to the inset of Fig. 2, the shift of the transition temperature determined by the shift of the peak of the calorimetric curves gives $|\Delta T|_{BCE} = 19.7$ K with $P = 2.69$ kbar. Yet a third way to indirectly estimate $\Delta T_{BCE}$ following ref. 7 is by using zero-field heat capacity data (inset of Fig. 4a) and the $\Delta S_t(T,P)$ curve (Fig. 2) giving $|\Delta T|_{BCE} = 16.2$ K for $P = 2.69$ kbar (as shown in Fig. 4a).

Just as the indirect methods likely overestimate $\Delta T_{BCE}$, direct measurements underestimate it due to non-adiabatic conditions that result in heat flow to and from the surroundings. In our case, direct decompression measurements ($P = 2.0$ kbar) resulted in $\Delta T_{BCE} \sim 3.1$ K (Fig. 4a). We have also designed a new experimental set-up that allowed us to measure the direct temperature change on *compression* using large sample masses (> 20 g). Compression-temperature measurements on a material of slightly different concentration ($x = 0.39$ rather than 0.38 so that the transition occurred near room temperature) resulted in $\Delta T_{BCE}$ values greater than $-4.3$ K for $P = 2.5$ kbar. These values of $\Delta T_{BCE}$ are comparable with the magnetic-field-induced adiabatic temperature changes for giant magnetocaloric materials for $\mu_0 H = 2$ T (see Table 1).

Because of the granule structure of (MnNiSi)$_{1-x}$(FeCoGe)$_x$, it is difficult to determine the magnetic-field-induced $\Delta T_{MCE}$ accurately by employing direct measurements, or indirectly using field-dependent heat capacity data. In the future, a noncontact method using a pulsed magnetic field could be implemented to determine $\Delta T_{MCE}$ as proposed in ref. 27. However, the shift in $T_M$ by 6 K for $\mu_0 H = 5$ T indicates that a $\Delta T_{MCE} \sim 2.4$ K could be expected for $\mu_0 H = 2$ T. As shown in Fig. 4b, an increase in the width of the working temperature range, i.e., an effective increase in the cooling efficiency, is possible by applying a magnetic field during depressurization and vice versa. In other words, the range increases if both pressure and field can be implemented in a cooling cycle, and the temperature changes due to the MCE and BCE may be added.

**Discussion**

The enhanced transition entropy with pressure (Fig. 2) is a consequence of an additional pressure-induced volume change (i.e., in addition to that which occurs as a result of the magnetostructural transition). The relationship between the structural entropy ($\Delta S_{st}$) and the volume change has been established previously as $\delta[\Delta V/V(\%)]/\delta(\Delta S_{st}) = 0.08$ J kg$^{-1}$ K$^{-1}$.[28] The excess entropy of about 11.9 J kg$^{-1}$ K$^{-1}$ for $P = 2.69$ kbar corresponds to a pressure-induced volume change of about 1%. The saturation magnetization at low temperature measured at different constant pressures remains almost unchanged (as shown in Fig. 1c). This suggests only a minor variation in the ferromagnetic exchange in the low temperature orthorhombic phase and rules out the possibility of exchange-enhanced magnetostriction. The deviation from linear Curie-Weiss behavior of the temperature-dependent inverse susceptibility (see Fig. 1d) suggests that the additional volume change is connected to a magneto-volume effect due to itinerant magnetic behavior[29] in the hexagonal paramagnetic state. This is not the usual type of volume change resulting from spontaneous magnetostriction in an itinerant magnetic system because the change in amplitude of the local moment at the transition is connected to the magnetocrystalline coupling responsible for the MST. The reduction of the disordered local moment in the paramagnetic state can also induce a volume change in an itinerant system, such as that observed in Invar alloys,[29,30] and recently reported in Mn$_3$GaN (ref. 7). A similar explanation fits in our case: instead of a spontaneous magnetostriction in the paramagnetic state, a pressure-induced magnetostriction is responsible for a volume change in addition to that caused by the structural transition, resulting in an additional change in entropy.

In the Mn*TX* compounds, a noncollinear magnetic state preferentially stabilizes through a reduction of the density of states at the Fermi level via a broadening of the 3*d* band.[31] Therefore, the application of pressure can promote a noncollinear magnetic state with a reduced magnetic moment.[23,31,32] However, in the paramagnetic state with short-range order, the application of pressure can result in an additional volume contraction associated with a forced magnetostriction based on the spin-fluctuation theory of itinerant systems.[24] Spin fluctuations prevent the formation of long-range noncollinear magnetic ordering through the magnetic energy gain due to short-range order. As a result, a volume contraction occurs due to band broadening to lower the electronic kinetic energy. Therefore, two consecutive effects, a magnetostructural transition followed by a pressure-induced magnetostriction, are responsible for the giant barocaloric effects observed in $(MnNiSi)_{1-x}(FeCoGe)_x$.

In summary, $(MnNiSi)_{1-x}(FeCoGe)_x$ ($x = 0.38$) exhibits both large baro- and magneto-caloric effects that exceed the best multicaloric materials known so far. The combined caloric effects could be useful to optimize the solid-state cooling efficiency desirable for practical applications. An additional volume change responsible for the pressure-induced enhancement of the barocaloric effect arises in a unique way: the volume change due to the magnetostructural transition transforms the system from a localized magnetically ordered state to an itinerant paramagnetic state. A decrease in amplitude of the disordered local moment in the spin fluctuating paramagnetic state, correlated with a pressure-induced magnetostriction, results in a concomitant contraction of the lattice associated with a magneto-volume effect in the itinerant magnetic system. We anticipate that the giant barocaloric effects observed in this new system will inspire the future development of multicaloric materials, the properties of which could be exploited to advance the development of solid-state-cooling.

**Methods**

Polycrystalline (MnNiSi)$_{1-x}$(FeCoGe)$_x$ ($x$ = 0.38 and 0.39) samples were prepared by melting the constituent elements of purity better than 99.9% in an ultra-high purity argon atmosphere using an RF-furnace. The samples were annealed under high vacuum for 3 days at 750°C followed by quenching in cold water. The phase purity of the samples were determined using a room temperature X-ray diffractometer (XRD) employing Cu $K\alpha$ radiation. A superconducting quantum interference device magnetometer (SQUID, Quantum Design MPMS) was used to measure the magnetization within the temperature interval of 10-380 K, and in applied magnetic fields up to 5 T. Magnetic measurements under hydrostatic pressure ($P$) were performed in a commercial BeCu cylindrical pressure cell (Quantum Design). Daphne 7373 oil was used as the pressure transmitting medium. The value of the applied pressure was calibrated by measuring the shift of the superconducting transition temperature of Pb used as a reference manometer (Pb has a critical temperature $T_C \sim$ 7.19 K at ambient pressure).[33] The isothermal entropy change ($\Delta S_{\text{MCE}}$) due to the magnetocaloric effect was estimated from the isothermal magnetization curves [$M(\mu_0 H)$] using the integrated Maxwell relation, $\Delta S_{\text{MCE}} = \int_0^{\mu_0 H} \left(\frac{\partial M}{\partial T}\right)_H \mu_0 dH$. Zero-field heat capacity measurement was performed using a Physical Properties Measurement System (PPMS by Quantum Design, Inc.) in a temperature range of 2–358 K. Calorimetric measurements with and without the application of hydrostatic pressures were carried out employing a purpose-built calorimeter as described in ref. 6. The transition entropy change (referenced to a given state at $T_0$) was estimated from the calorimetric curves using the relation

$$\Delta S_t(T,P) = S(T,P) - S(T_0,P) = \int_{T_0}^{T} \frac{1}{T}\frac{\dot{Q}(P)}{\dot{T}}\,dT$$

where $\dot{Q}(P)$ and $\dot{T}$ are the heat flux and heat rate, respectively. Direct measurements of the adiabatic temperature change on the fast release of hydrostatic pressure were performed following the procedure described in ref. 12. The adiabatic temperature change during pressurization of the sample was carried out using an another purpose-built set-up. A type-J thermocouple was embedded in the mixture of powdered sample and a pressure transfer medium (a mixture of methanol and ethanol with a ratio 4:1). The sample was pressurized to > 2 kbar in about 30 s, and the temperature data were recorded every 0.04 s. The adiabatic temperature change estimated from zero-field heat capacity and the $\Delta S_t(T,P)$ curve was calculated following ref. 7.

**Acknowledgements**

Work at Louisiana State University (S.S.) was supported by the U.S. Department of Energy (DOE), Office of Science, Basic Energy Sciences (BES) under Award No. DE-FG02-13ER46946, and heat capacity measurements were carried out at LSU by P.W.A. who is supported by DOE, Office of Science, BES under Award No. DE-FG02-07ER46420. The calorimetric measurement under applied hydrostatic pressure was supported by the CICYT (Spain), under grant FIS2014-54734-P and by the Catalan Government (Grant 2014SGR-00581). P.L. acknowledges support from SUR (DEC,Catalonia). Work at Southern Illinois University was supported by DOE, Office of Science, BES under Award No. DE-FG02-06ER46291. D.P.Y. acknowledges support from the NSF through DMR Grant No. 1306392.

**Author contributions**

T.S. and S.S. designed the experiment, and wrote the manuscript. P.L., M.B. and J.Ll.T. carried out calorimetric measurements under hydrostatic pressure. S.S. A.U.S. and T.S. designed the experimental set-up for compression measurements and performed the experiment to determine the adiabatic temperature change. P.W.A. carried out heat capacity measurement. D.P.Y. and T.S. fabricated the samples. D.L. E.K. and T.S. prepared samples and conducted magnetic measurement. N.A. and I.D. performed X-ray diffraction measurement and carried out preliminary magnetic characterization. All authors reviewed the manuscript.

**Additional information**

**Competing financial interests:** The authors declare no competing financial interests.

**Table 1 | Giant multicaloric effects at first-order phase transitions.**

| Materials | $T$ | $|\Delta S|_{BCE}$ | $|\Delta T|_{BCE}$ | RCP | $P$ | $|\Delta S|_{MCE}$ | $|\Delta T|_{MCE}$ | RCP | $\mu_0 H$ | $\rho$ | Ref. |
|---|---|---|---|---|---|---|---|---|---|---|---|
| | K | mJ cm$^{-3}$ K$^{-1}$ | K | J cm$^{-3}$ | kbar | mJ cm$^{-3}$ K$^{-1}$ | K | J cm$^{-3}$ | T | g cm$^{-3}$ | |
| Ni$_{49.26}$Mn$_{36.08}$In$_{14.66}$ | 293 | *200* | [4.5] | 1 | 2.6 | 82 | [1.3] | 0.3 | 0.94 | 8.2 | 6 |
| Gd$_5$Si$_2$Ge$_2$ | 270 | *82.5* | {1.1}* | 1 | 2.9* | 120 | [7] | 0.96 | 2 | 7.5 | 1,12 |
| LaFe$_{11.33}$Co$_{0.47}$Si$_{1.2}$ | 237 | *63.5* | {2.2}* | 1.3 | 2.0* | 76 | {0.9} | 0.33 | 5 | 7.3 | 13 |
| Fe$_{49}$Rh$_{51}$ | 308 | *123.5* | [8.1] | 1 | 1.1 | 120 | [6] | 2.5 | 2 | 9.8 | 24,25 |
| MnCoGe$_{0.99}$In$_{0.01}$ | 308 | *413* | [18.5] {9.4}$^{\#}$ | 9.5 | 3.0 3.0$^{\#}$ | 95 | [2.8] | 0.63 | 2 | 7.95 | 10,15 |
| Mn$_3$GaN | 285 | *170* | [4.8] | 1.5 | 1.39 | – | – | – | – | 7.6 | 7 |
| (NH$_4$)$_2$SO$_4$ | 219 | *106* | [8] | 0.5 | 1.0 | – | – | – | – | 1.8 | 8 |
| (MnNiSi)$_{0.62}$(FeCoGe)$_{0.38}$ | 338 | *538* | [33] {3.1}* {4.3}$^{\dagger}$ | 11 | 2.69 2.0* 2.5$^{\dagger}$ | 152 | [2.4] | 0.9 | 2 | 7.3 | This work |

Isothermal entropy change $|\Delta S|$, adiabatic temperature change $|\Delta T|$, and relative cooling power RCP, due to changes of hydrostatic pressure $P$ (barocaloric, BCE) and magnetic field $H$ (magnetocaloric, MCE). Entries inside curly brackets {…} were derived from direct measurements. *Data acquired from hydrostatic decompression measurements. $^{\dagger}$Data acquired from hydrostatic compression measurement using a slightly different concentration $x$ = 0.39. $^{\#}$Data derived from direct measurement by pressurizing the sample using hydraulic press. Entries inside square brackets […] denote parameters derived from $-c\Delta T \approx T\Delta S$ using zero-field heat capacity data. Entries in italic font signify data derived from quasi-direct[5] measurements. No data available for the dashed entries. $\rho$ is the mass density.

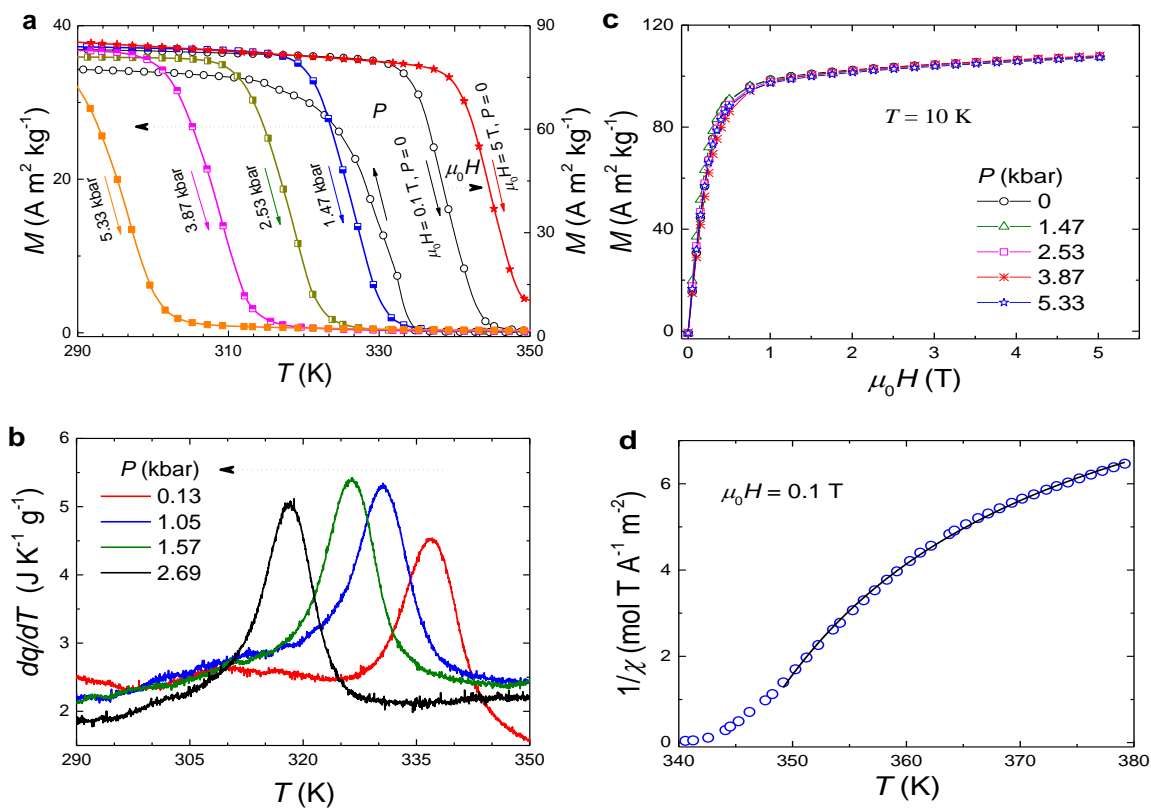

**Figure 1 | Magnetostructural phase transitions under applied magnetic field and pressure. a**, The magnetization ($M$) of $(MnNiSi)_{1-x}(FeCoGe)_x$ ($x = 0.38$) as a function of temperature ($T$) showing the shifts in the transition temperature with pressure ($P$) and magnetic field ($\mu_0 H$). The dotted arrows indicate the shifts of the transition with pressure and magnetic field, respectively. **b**, Calorimetric heat flow curves ($dq/dT$) for selected values of hydrostatic pressures. **c**, Field-dependent magnetization at $T = 10$ K at different hydrostatic pressures. **d**, The inverse susceptibility as a function of temperature for $\mu_0 H = 0.1$ T, and the corresponding fitting (solid curve) using a modified Curie-Weiss expression, $\chi = \chi_0 + C/(T - \Theta_P)$, where $\Theta_P$ is the paramagnetic Curie temperature. The nonlinear behavior indicates itinerant magnetism.

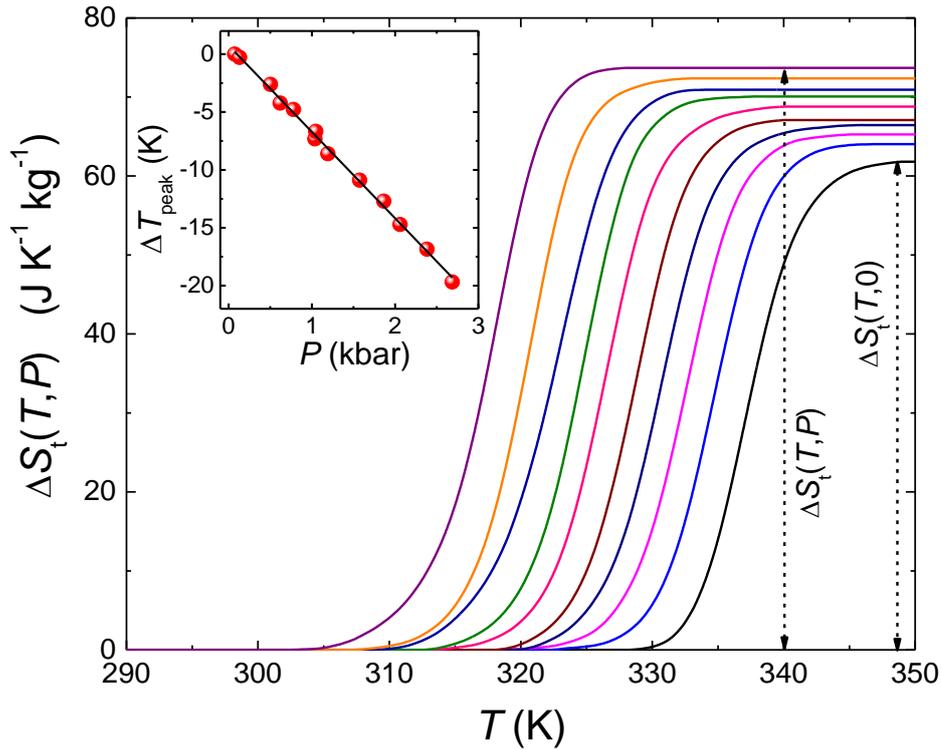

**Figure 2 | Pressure dependent transition entropy as a function of temperature.**
Entropy change ($\Delta S_t$) relative to the low-temperature phase for selected applied hydrostatic pressures ($P$). From left to right, the curves correspond to 2.69, 2.38, 2.06, 1.86, 1.57, 1.19, 1.05, 0.78, 0.50, and 0.07 kbar. The pressure-induced entropy increases from $\Delta S_t(T,0)$ to $\Delta S_t(T,P)$ as $P$ goes from 0 to 2.69 kbar. The inset shows the temperature shifts in the peaks of the calorimetric curves as a function of applied pressure.

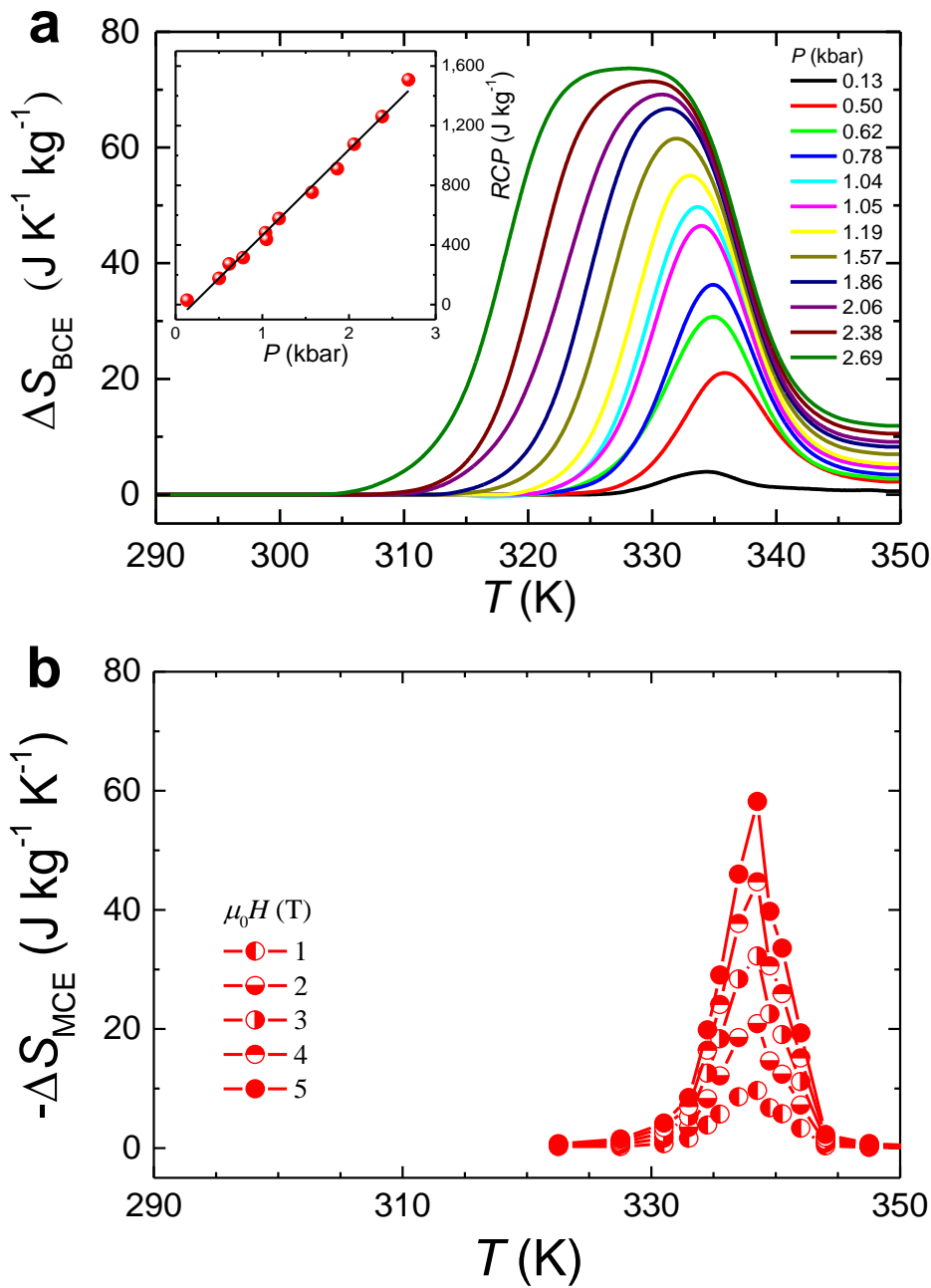

**Figure 3 | Giant multicaloric effects in (MnNiSi)$_{1-x}$(FeCoGe)$_x$. a,b,** Entropy changes associated with (**a**) pressure-induced (inverse) barocaloric and (**b**) magnetic-field-induced (conventional) magnetocaloric effects with the isothermal application of pressures up to 2.69 kbar and magnetic fields up to 5 T, respectively. The inset of (**a**) shows the relative cooling power (RCP = |$\Delta S$|$_{BCE}$×$\delta T_{FWHM}$) as a function of pressure.

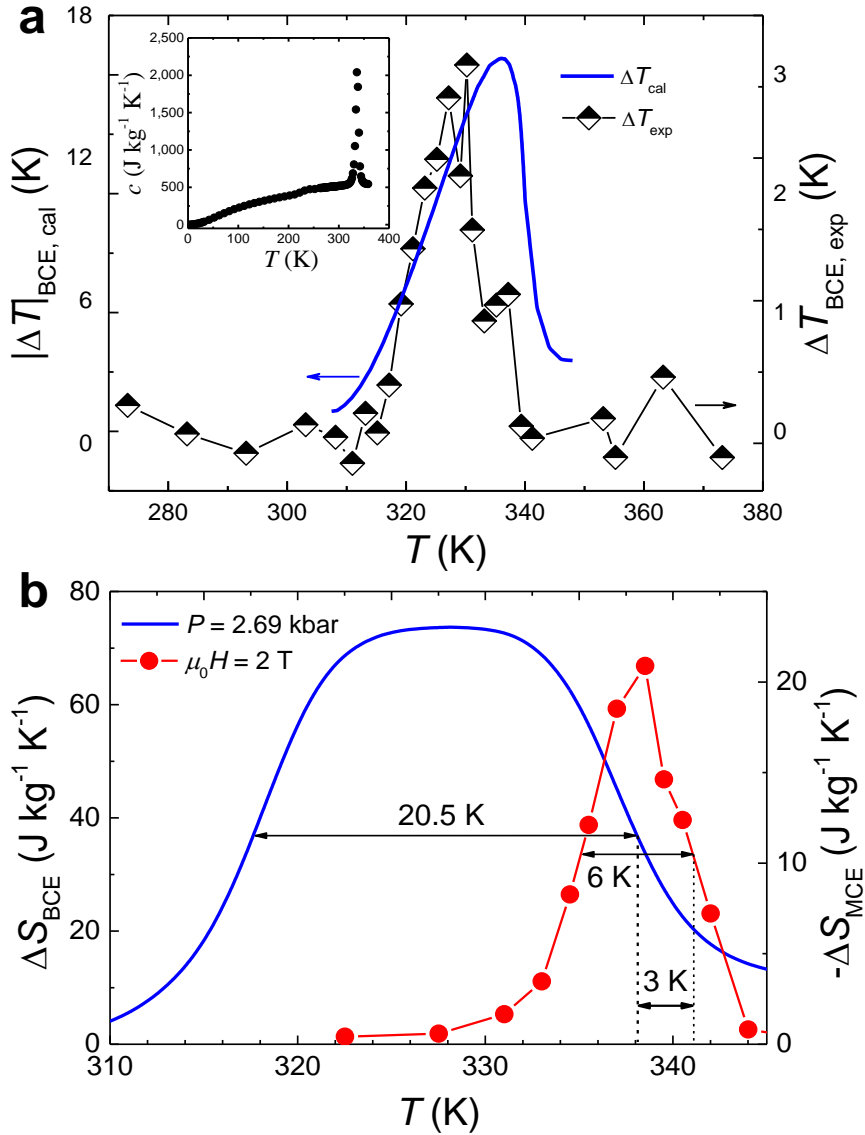

**Figure 4 | Adiabatic temperature change in (MnNiSi)$_{1-x}$(FeCoGe)$_x$. a**, The adiabatic temperature change upon the depressurization from 2 kbar to atmosphere (right axis), and the same estimated from zero-field heat capacity data (left axis). The heat capacity as a function of temperature is shown in the inset. **b**, The pressure-induced entropy change for $P$ = 2.69 kbar and the magnetic-field-induced change for $\mu_0H$ = 2 T overlaid to illustrate the potential extension of the width of the entropy versus temperature curve that might be realized by a cooling cycle that can employ both hydrostatic pressure and magnetic field.

20